\documentclass[prd,twocolumn,showpacs,graphicx,print]{revtex4-1}
\usepackage{graphicx}
\usepackage{dcolumn}
\usepackage{bm}
\usepackage{epsfig}
\usepackage{lipsum}
\usepackage{natbib}
\usepackage{ulem,cancel,soul}
\usepackage[fleqn]{amsmath}
\usepackage{slashed}
\usepackage{mathrsfs}
\usepackage{amssymb}
\usepackage{times}
\usepackage{psfrag}
\usepackage[english]{babel}
\usepackage{multirow}
\usepackage{color}
\begin{document}

\title{Quantum Gravity Correction to Dirac Equation via Vacuum Coupling Contribution}

\author{Sijo K. Joseph}
\affiliation{Quantum Gravity Research, Topanga, CA 90290, USA.}
\email{sijo@quantumgravityresearch.org}

\begin{abstract}
Generalized Dirac equation containing vacuum-mass contribution is introduced. The vacuum-mass contribution arises due to the coupling of quantum mechanical matter field with the vacuum field. Vacuum stress energy tensor arises in the Scalar-Tensor Field equation (Extended Einstein's Field equation), implies a mass like corrections to the newly formulated Dirac equation. Apart from the inertial mass of the electron, generalized  Dirac equation proposed here, contains a mass-term as a result of vacuum field coupling.
\end{abstract}

\date{\today}


\maketitle

\section{Introduction}
Geometrical interpretations of quantum mechanics have attracted much attention in recent years~\cite{Shojai_Article,QMGeometry,GabayJoseph1,GabayJoseph2}.
Dualities between quantum field theory and general relativity must results from the more fundamental fact that quantum mechanics can be formulated purely in-terms of geometry. Extending Maldacena's ER=EPR towards  Susskind's  more general GR=QM  proposal is strongly suggesting that quantum theory must be entirely based on geometry. One can also put-forward a curious question about the nature of the geometry of quantum theory. Nottale had pointed out that the quantum theory naturally arises from the the most general geometry of nature, the fractal geometry \cite{NottaleBook1,NottaleBook2}. Due to the non-differentiable nature of
fractals, it is hard to find an extended version of differential geometry incorporating fractal manifolds.
One can use an infinite sum of smooth manifolds to approximate a fractal manifold.
It is already known that quantum-potential is associated to the fractality of the space-time~\cite{NottaleBook1,NottaleBook2}. In a more simplified scenario, one can take into account the scale symmetry of fractals on a differential manifold to incorporate the quantum mechanical nature of matter field and it will appear as an effective theory. The fractal contributions can appear as a small correction to the smooth Riemannian manifold, hence Einstein's Gravitational field contribution must contain small quantum mechanical corrections due to the micro-curvature or fractality of space-time; or in other words the conformal fluctuation of the metric. 
In order to incorporate these micro-curvatures, one needs to look into extended versions of Einstein's theory~\cite{BeyondEinsteinGravity,WeylReview2017,Bergmann1968,fR_HamFormlation,fofT1,*fofT2,Bekenstein2011_TeVeS,
*BekensteinPRD_TeVeS,Moffat_STVG,*MOG_Moffat,Shojai_Article}.
Using the deBroglie-Bohm version of Quantum theory~\cite{Bohm1975,BohmI,BohmII,Bohm1975,PHollandBook,SheldonReview}, one can combine Quantum Theory with General Theory of Relativity and reaches to a special class of Scalar-Tensor Theory which is explored in ~\cite{Shojai_Article,ShojaiCnstrAlgebra,ShojaiBohmianQM,Shojai_ScalarTensor,GabayJoseph1,GabayJoseph2} and this special class of Scalar-Tensor Theory is termed as Bohmian-Quantum Gravity. It is to be noted that, aforementioned theory is just a classical field theory in its geometric form; space-time is not quantized like in Canonical Quantum Gravity
~\cite{Thiemann_2007} or like in Loop Quantum Gravity~\cite{rovelli_vidotto_2014,Gambini2011,AshtekarBook2017}, 
gravity is still treated as classical; space-time is continuous but the quantum nature of matter field on a smooth manifold is the main aspect of the theory. In the continuum limit of a sucessfull Quantum-Gravity theory, one should reproduce Bohmian Quantum Gravity kind of theory. Exploring such an interesting theory reveals an unexplored fluctuating backround field $\lambda$ which appears as the Lagrage multiplier in the theory. It is already shown that quantum mechanical uncertainty relation can be deduced from this fluctuating field $\lambda$ ~\cite{GabayJoseph1}. It was Nelson who proposed a stochastic mechanics approach to quantum mechanics \cite{NelsonBook}. In his approach, quantum particles are driven by a kind of Browninan motion resulting from quantum fluctuation. Physical orgin of such a fluctuation is still mysterious. It was S. Roy who suggested that the random zero point field induces the probabilistic aspect in the geometry of background space-time~\cite{Sisir_Roy1}. G. S. Asanov, S. P. Ponomarenko and S. Roy derived four dimensional quantum mechical conformal factor using five-dimensional probabilistic Finsler metric~\cite{Sisir_Roy2}. According to them,  any microscopic system will have stochastic behavior due to the presence of fluctuating vacuum characterized by the probabilistic Finsler metric. According to them, 
underlying Finsler manifold plays a great role to understand the quantum behavior in a consistent and complete way.
The nature of quantum fluctuations and its possible relationship with the gravitational field is also explored by L. Smolin~\cite{Smolin_QFluct1}. Bohmian-Quantum gravity also hints towards the idea that quantum particle might be immersed in some kind of fluctuating vacuum field. 

In Bohmian-Quantum Gravity, such a fluctuating vacuum field  appears from the unification of gravitation and quantum mechanics  using deBroglie-Bohm version of Quantum theory~\cite{Bohm1975,BohmI,BohmII,PHollandBook,SheldonReview}. For such a unification, one needs to take three crucial steps, (a) Extend Einstein's Tensor Theory to Scalar-Tensor Theory (b) Accept the generalization of deBroglie-Bohm picture of Quantum theory (Note that deBroglie-Bohm theory is only linear order in quantum potential $Q$) (c) Accept the presence of a fluctuating vacuum field which appears as the Lagrange multiplier in the theory.
Taking all these factors into account the following action is proposed ~\cite{GabayJoseph1,GabayJoseph2}, 
\begin{eqnarray}
& & A[g_{\mu\nu},{\Omega}, S, \rho, \lambda]=\nonumber\\
& & \frac{1}{2\kappa}\int{d^4x\sqrt{-g}\left(R\Omega^2-6\nabla_{\mu}\Omega\nabla^{\mu}\Omega\right)}  \nonumber \\
& & +\int{d^4x\sqrt{-g} \left(\frac{\rho}{m}\Omega^2 \nabla_{\mu}S \nabla^{\mu}S-m\rho\Omega^4\right)} \nonumber \\
& & +\int{d^4x\sqrt{-g}\lambda\left[\ln{\Omega^2}-\left(\frac{\hbar^2}{m^2}\frac{\nabla_{\mu}\nabla^{\mu}\sqrt{\rho}}{\sqrt{\rho}}
\right)\right]} \label{Actioneq}. 
\end{eqnarray}
and Lagrange multiplier $\lambda$ is identified as a scalar vacuum field. 
It is already shown that, minimizing the action with respect to $\rho$ and $S$ 
leads to real and imaginary parts of the Generalized Klein-Gordon Equation. 
It can be easily seen that the real part of the Klein-Gordon Equation obeys the following equation 
\begin{equation}
\begin{split}
\nabla_{\mu}S \nabla^{\mu}S-m^2\Omega^2+\frac{\hbar^2}{2m 
\Omega^2\sqrt{\rho}}\Bigl[\Box{\Bigl(\frac{\lambda}{\sqrt{\rho}}\Bigr)}-\lambda\frac{\Box\sqrt{\rho}}{\rho}\Bigr]=0. \label{EqMotion} 
\end{split}
\end{equation}
while the imaginary part gives the generalized continuity equation
\begin{eqnarray}
\nabla_{\mu}(\rho\Omega^2\nabla^{\mu}S)=0 \label{ContiEqn}.
\end{eqnarray}
Combining these two equations (Eq.~\ref{EqMotion} and Eq.~\ref{ContiEqn}) into a single complex wave equation, the generalized Klein-Gordon Equation can be obtained.

   
\section{Modified Klein-Gordon Equation to Modified Dirac Equation}
Bohmian equation of motion and continuity equation (Eq.~\ref{EqMotion} and Eq.~\ref{ContiEqn}), without vacuum field contributions (taking $\lambda=0$), can be expressed in terms of the density ${\rho}$, classical action $S$, and conformal factor $\Omega^2$ as already explored in the previous works~\cite{Shojai_Article,QMGeometry,GabayJoseph1,GabayJoseph2},
\begin{eqnarray}
\nabla_{\mu}S\nabla^{\mu}S - m^2\Omega^2=0 \label{EoMnoVac}\\
\nabla_{\mu}(\rho\Omega^2\nabla^{\mu}S)=0 \label{EoMcnteqn}.
\end{eqnarray}
Here, $\Omega^2=\exp(\frac{\hbar^2}{m^2} \frac{\nabla_{\mu} \nabla^{\mu}\sqrt{\rho}}{\sqrt{\rho}})=\exp(Q)$, where $Q=\frac{\hbar^2}{m^2} \frac{\nabla_{\mu} \nabla^{\mu}\sqrt{\rho}}{\sqrt{\rho}}$, is the Klein-Gordon quantum potential. 

One can combine Eq.~\ref{EoMnoVac} and Eq.~\ref{EoMcnteqn} in order to obtain the modified Klein-Gordon Equation. Hence the generalized Klein-Gordon equation is given by,
\begin{eqnarray}
\Box\phi+\frac{1}{2}\frac{\nabla_{\mu}\Omega^2}{\Omega^2}\Bigl(\frac{\nabla^{\mu}\phi}{\phi}-\frac{\nabla^{\mu}\phi^{*}}{\phi^{*}}\Bigr)\phi+ \frac{m^2}{\hbar^2}\phi=0 \label{WFeqnp1} 
\end{eqnarray}
Using the fact that,
\begin{eqnarray}
\frac{1}{2}\Bigl(\frac{\nabla^{\mu}\phi}{\phi}-\frac{\nabla^{\mu}\phi^{*}}{\phi^{*}}\Bigr)=\frac{i}{\hbar}\nabla^{\mu}S,\label{SDer}
\end{eqnarray}
Eq.~\ref{WFeqnp1} can be written as,
\begin{eqnarray}
\Box\phi+\frac{i}{\hbar}\Bigl(\frac{\nabla_{\mu}\Omega^2}{\Omega^2}{\nabla^{\mu}S}\Bigr)\phi+ \frac{m^2}{\hbar^2}\phi=0. 
\label{WFeqn1} 
\end{eqnarray}
The expression $\frac{\nabla_{\mu}\Omega^2}{\Omega^2}$ can be perceived as the quantum force for an exponential constraint 
($\Omega^2=e^{Q}$) of the conformal factor $\frac{\nabla_{\mu}\Omega^2}{\Omega^2}= \nabla_{\mu}\ln{\Omega^2}=\nabla_{\mu}Q$. Hence,

\begin{eqnarray}
\Box\phi+\frac{i}{\hbar}\Bigl({\nabla_{\mu}Q\,\nabla^{\mu}S}\Bigr)\phi+ \frac{m^2}{\hbar^2}\phi=0. 
\label{WFeqn1} 
\end{eqnarray}

It is pretty straight forward to see that Eq.~\ref{EoMnoVac} and Eq.~\ref{EoMcnteqn} can give Eq.~\ref{WFeqn1}. In the usual Klein-Gordon equation, the continuity equation is given by $\nabla_{\mu}(\rho\nabla^{\mu}S)=0$ while the conformal gravity action (see Eq.~\ref{Actioneq})  generalizes the usual Klein-Gordon continuity equation $\nabla_{\mu}(\rho\nabla^{\mu}S)=0$ to $\nabla_{\mu}(\rho\Omega^2\nabla^{\mu}S)=0$ with an extra factor $\Omega^2$.
Continuity equation is usually derived from the imaginary part of the Klein-Gordon equation, hence this extra contribution should appear either as the part of $\nabla_{\mu}\phi$ or with a complex number times $\phi$.  Since the equation of motion (Eq.~\ref{EoMnoVac}) didn't get corrected by the conformal gravity, $\Box\psi+\frac{m^2}{\hbar^2}\psi$ remains the same. 
Due to the extra conformal factor $\Omega^2$ in the continuity equation, an additional imaginary term appears as the conformal gravity correction to the Klein-Gordon equation. One can see that the  middle term $+\frac{i}{\hbar}\Bigl({\nabla_{\mu}Q\,\nabla^{\mu}S}\Bigr)\phi$ arises as a gravitational correction to the Klein-Gordon equation and it is interesting to note that such a correction appears as a dissipative contribution to the wave-equation.

Here the aim is to further simplify Eq.~\ref{WFeqn1}, for that purpose, one need to have a good understanding about the operator nature of the quantum potential $Q$. It is a matter of switching between quantum density formalism (Bohmian formalism ) and wave-function  formalism. After some simple algebra involving $\phi$ and $\sqrt{\rho}$, one finds an interesting relationship between quantum potential $Q$ and scalar field $\phi$. Starting from the very basic relation, we get,
\begin{eqnarray}
\phi&=&\sqrt{\rho}\,\exp{\Bigr(\frac{i}{\hbar}S\Bigl)}\\
\nabla_{\mu}\phi&=&\nabla_{\mu}\sqrt{\rho}\,\exp{\Bigr(\frac{i}{\hbar}S\Bigl)}+\frac{i}{\hbar}\nabla_{\mu}S\,\phi\\
\Bigr(\nabla_{\mu}-\frac{i}{\hbar}\nabla_{\mu}S\Bigl) \phi&=&\nabla_{\mu}\sqrt{\rho}\,\exp{\Bigr(\frac{i}{\hbar}S\Bigl)}.
\end{eqnarray}
As a more general statement, it can be proven that 
$f(\nabla_{\mu})\phi=f\Bigr(\overset{+}{\mathcal{D}}_{\mu}\sqrt{\rho}\Bigl)\exp{\Bigr(\frac{i}{\hbar}S\Bigl)}$ where $\overset{+}{\mathcal{D}}_{\mu}=(\nabla_{\mu}+\frac{i}{\hbar}\nabla_{\mu}S)$. One can also define ${\mathcal{D}}_{\mu}=(\nabla_{\mu}-\frac{i}{\hbar}\nabla_{\mu}S)$ and the following operator relationships are obtained,

\begin{eqnarray}
\frac{f(\overset{+}{\mathcal{D}}_{\mu})\sqrt{\rho}}{\sqrt{\rho}}=\frac{f(\nabla_{\mu})\phi}{\phi}\label{DplusEq}
\end{eqnarray}

\begin{eqnarray}
\frac{f(\nabla_{\mu})\sqrt{\rho}}{\sqrt{\rho}}=\frac{f({\mathcal{D}}_{\mu})\phi}{\phi}\label{DminusEq}.
\end{eqnarray}

These operator relationships are really useful to switch between the quantum mechanical density $\rho$ and the wave-function $\phi$ formalism. Ignoring $\frac{\hbar^2}{m^2}$ factor from the quantum potential, define a new quantity $\widetilde{Q}=\frac{\Box{\sqrt{\rho}}}{\sqrt{\rho}}$, one can
explore the operator correspondence. 
Rearranging Eq.~\ref{DminusEq} and considering operator $\nabla_{\mu}\nabla^{\mu}$,  Eq.~\ref{DminusEq3} can be deduced,
\begin{eqnarray}
\Bigl(\frac{f(\nabla_{\mu}\nabla^{\mu})\sqrt{\rho}}{\sqrt{\rho}}\Bigl)\phi=f({\mathcal{D}}_{\mu}{\mathcal{D}}^{\mu})\phi\label{DminusEq3}.
\end{eqnarray}
Equation~\ref{DminusEq3} is just another statement of the exponential shift theorem for the linear differential operators $f(D)(e^{ax}\sqrt{\rho})\equiv e^{ax}f(D+a)\sqrt{\rho}$.
Applying Eq.~\ref{DminusEq3}, the quantity $\widetilde{Q}\phi$ can be seen as an operator acting on $\phi$. Note that, in the Bohmian perspective quantum potential $Q$ is just a scalar function of $\sqrt{\rho}$. 

\begin{eqnarray}
\widetilde{Q}\phi=\Bigr(\frac{\Box{\sqrt{\rho}}}{\sqrt{\rho}}\Bigl)\phi={\mathcal{D}}_{\mu}{\mathcal{D}}^{\mu}\phi\label{Qpsi}
\end{eqnarray}

After a straight forward expansion of the expression $\widetilde{Q}\phi=0$, it is found that the expression is equivalent to Eq.~\ref{WFeqn1}, once the classical energy momentum relation 
$\nabla_{\mu}S\nabla^{\mu}S=m^2$ is assumed. Hence the equation can be written in a more compact form,

\begin{eqnarray}
\Box\phi+\frac{i}{\hbar}\Bigr(\nabla_{\mu}Q\nabla^{\mu}S\Bigl)\phi+ \frac{m^2}{\hbar^2}\phi=0 \implies {\mathcal{D}}_{\mu}{\mathcal{D}}^{\mu}\phi=0 \nonumber\\ \label{DisipKGEq} 
\end{eqnarray}

This quantum gravity modified generalized Klein-Gordon equation, looks like a massless equation in terms of the gauge covariant derivative ${\mathcal{D}}_{\mu}$.
\begin{eqnarray}
{\mathcal{D}}_{\mu}{\mathcal{D}}^{\mu}\phi=0, \label{KleinNice}
\end{eqnarray}
where ${\mathcal{D}}_{\mu}=(\nabla_{\mu}-\frac{i}{\hbar}\nabla_{\mu}S)$.
Finding the Dirac equation from Eq.~\ref{KleinNice} is straight forward and it is given by, 
\begin{eqnarray}
 i\gamma^{\mu}{\mathcal{D}}_{\mu}\Psi=0, \label{DiracNice}
\end{eqnarray}
Expanding ${\mathcal{D}}_{\mu}$ operator, the following equation is obtained,

\begin{eqnarray}
(i\gamma^{\mu}{\nabla}_{\mu}+\frac{1}{\hbar}\gamma^{\mu}\nabla_{\mu}S)\Psi=0, \label{DiracNice}
\end{eqnarray}

where $\gamma^{\mu}$ describe Dirac matrices and $\Psi$ is a spinor field.
One can assign $\gamma^{\mu}\nabla_{\mu}S=-m I_{4}$ to match it with the Dirac equation
\begin{eqnarray}
(i\hbar\gamma^{\mu}{\nabla}_{\mu}-m)\Psi=0. \label{DiracNice2}
\end{eqnarray}
One can make a curious observation that $\nabla_{\mu}S$ looks like electro-magnetic gauge connection $A_{\mu}$, if $\nabla_{\mu}S\propto e A_{\mu}$, even the rest mass $m$ can be seen as an electro-magnetic quantity $m \propto e\gamma^{\mu}A_{\mu}$.
Note that,  in the Dirac equation given in Eq.\ref{DiracNice2}, rest mass $m$ is a function of space-time variables, hence $\nabla_{\mu}m$ can give an imaginary contribution while squaring the Dirac equation and it should match with the imaginary part appearing the generalized Klein-Gordon equation given in Eq.~\ref{DisipKGEq}. One need to be aware that, squaring Eq.\ref{DiracNice2} should be  done without using the conjugate of the complex differential ${\mathcal{D}}_{\mu}$ to re-obtain the Klein-Gordon Equation given in Eq.\ref{KleinNice}.  

In order to get the complete form of Eq.\ref{KleinNice}, incorporating an extra mass term, one needs to take into account the vacuum field $\lambda$ contribution. So far we have dealt with the action $A$, while ignoring the $\lambda$
contribution (Eq.~\ref{Actioneq} with $\lambda=0$). Considering only the conformal gravity correction a dissipative term appears in Eq.~\ref{DisipKGEq}, in order to balance the dissipation, one must incorporate a forcing contribution in the equation. This contribution naturally appears when we take into account the quantum vacuum contribution arising from $\lambda$ field.


\section{Generalized Klein-Gordon with extra mass via Lagrange multiplier}
Taking into account the $\lambda$ contribution, as in the previous case, two real field equation can be obtained
\begin{eqnarray}
& & \nabla_{\mu}S \nabla^{\mu}S-m^2\Omega^2+\frac{\hbar^2}{2m 
	\Omega^2\sqrt{\rho}}\Bigl[\Box{\Bigl(\frac{\lambda}{\sqrt{\rho}}\Bigr)}-\lambda\frac{\Box\sqrt{\rho}}{\rho}\Bigr]=0. \nonumber \\ \label{EqMotionGen} 
\\& & \nabla_{\mu}(\rho\Omega^2\nabla^{\mu}S)=0 .\label{ContiEqGen}
\end{eqnarray}

Wave-function equation can be defined in the same manner as before. After substituting Eq.~\ref{ContiEqGen} into Eq.~\ref{EqMotionGen}, more general wave-function equation coupled to the quantum vacuum can be obtained, 
\begin{eqnarray}
\mathcal{D}_{\mu}\mathcal{D}^{\mu}\phi 
- \Bigl[\frac{1}{2m\Omega^2\rho}\Bigl(\Box
- \frac{2m^2(1-Q)}{\hbar^2}\Bigr)\lambda\Bigr]\phi =0 \label{WFeqnVacuumF1}. 
\end{eqnarray}
The vacuum density $\lambda$ obeys a first order differential equation coupled to  the density via $\sqrt{\rho}$ 
\begin{eqnarray}
\lambda=\frac{\hbar^2}{m^2(1-Q)}\nabla_{\mu}\Bigl(\lambda\frac{\nabla^{\mu}\sqrt{\rho}}{\sqrt{\rho}} \Bigr) \label{LambdaNiceEq1}
\end{eqnarray}
Eq.~\ref{WFeqnVacuumF1} and Eq.~\ref{LambdaNiceEq1} together yields, quantum gravity corrected Klein-Gordon equation with vacuum contribution.
From Eq.~\ref{WFeqnVacuumF1} and Eq.~\ref{LambdaNiceEq1}, it can be seen that $\lambda$ gives a mass like contribution to the equation.
Defining a vacuum mass contribution $M(\lambda,\rho)$ in the following way,
\begin{eqnarray}
 \Bigl[\frac{1}{2m\Omega^2\rho}\Bigl(\Box
- \frac{2m^2(1-Q)}{\hbar^2}\Bigr)\lambda\Bigr]\phi= -\frac{M(\lambda,\rho)^2}{\hbar^2}\phi \label{WFeqnVacuumF} ;
\end{eqnarray}

Modified Klein-Gordon equation can be written as,
\begin{eqnarray}
\Bigr(\mathcal{D}_{\mu}\mathcal{D}^{\mu}+\frac{M(\lambda,\rho)^2}{\hbar^2}\Bigl)\Phi=0 \label{WFeqnVacuumF}. 
\end{eqnarray}

Here the gravitationally corrected Klein-Gordon equation written in terms of ${\mathcal{D}}_{\mu}=(\nabla_{\mu}-\frac{i}{\hbar}\nabla_{\mu}S)$ looks like the usual Klein-Gordon equation $(\Box+m^2/\hbar^2)\Phi=0$. 

\section{Vacuum corrections to Dirac equation}

Its is already shown that the Quantum-Gravity Modified Klein-Gordon equation can be written as,
\begin{eqnarray}
\Bigr(\mathcal{D}_{\mu}\mathcal{D}^{\mu}+\frac{M(\lambda,\rho)^2}{\hbar^2}\Bigl)\Phi=0  \label{GenKGM}. 
\end{eqnarray}

Now applying the Dirac's technique of taking the square root of the operator $\Bigr(\mathcal{D}_{\mu}\mathcal{D}^{\mu}+M(\lambda,\rho)^2\Bigl)$, the corresponding gravity corrected fermionic equation can easily be found. Corresponding modified 
Dirac equation is given by,

\begin{eqnarray}
\Bigl(i\hbar\gamma^{\mu}\mathcal{D}_{\mu}-M(\lambda,\rho)\Bigr)\Psi=0 \label{WDiracGen}. 
\end{eqnarray}

Expressing above equation in terms of the usual space-time curvilinear ordinates $\nabla_{\mu}$,

\begin{eqnarray}
\Bigl(i\hbar\gamma^{\mu}\nabla_{\mu}+\gamma^{\mu}\nabla_{\mu}S- M(\lambda,\rho)\Bigr)\Psi=0 \label{WFeqnVacuumF}. 
\end{eqnarray}

Taking $\gamma^{\mu}\nabla_{\mu}S=-m I_{4}$, the usual Dirac equation is obtained with an extra vacuum mass contribution $M(\lambda,\rho)$,

\begin{eqnarray}
\Bigl(i\hbar\gamma^{\mu}\nabla_{\mu}-m-M(\lambda,\rho)\Bigr)\Psi=0 \label{WFeqnVacuumF}. 
\end{eqnarray}

\begin{eqnarray}
\lambda=\frac{\hbar^2}{m^2(1-Q)}\nabla_{\mu}\Bigl(\lambda\frac{\nabla^{\mu}\sqrt{\rho}}{\sqrt{\rho}} \Bigr) \label{LambdaNicePsibarEq}
\end{eqnarray}
Here $\Psi$ is the usual Dirac Spinor, $M(\lambda,\rho)= \pm\Bigl[\Bigl(\frac{m(1-Q)\lambda}{\rho}-\frac{\hbar^2}{2m\rho}\Box\lambda\Bigr)\Bigr]^{1/2}$ and $\rho=\bar{\Psi}\Psi$. Here vacuum contribution should obey an extra condition $\nabla_{\mu}M(\lambda,\rho)=0$, to match it with the generalized Klein-Gordon equation (Eq.~\ref{GenKGM}).
It is important to notice that Eq.~\ref{LambdaNicePsibarEq} is obtained from the Scalar-Tensor Einstein equation 
and the scalar curvature equation, for a detailed derivation see Ref.~\cite{GabayJoseph1,GabayJoseph2}.
In the case of the mass of the particle or antiparticle, $M(\lambda,\rho)$ can take positive or negative values (it can also be complex). Hence there are four ways to combine the rest mass of the particle with vacuum mass $M(\lambda,\rho)$ contribution. It is to be noted that the vacuum field may affect the particle and anti-particle differently. In order to get the Klein-Gordon equation one need to assign equal but opposite vacuum mass contributions $M(\lambda,\rho)$ to particle and antiparticle respectively. 

\begin{eqnarray}
\bar{\Psi}\Bigl(i\hbar\gamma^{\mu}\nabla_{\mu}+m+M(\lambda,\rho)\Bigr)=0 \label{WFeqnVacEplus} 
\end{eqnarray}

There exist another set of Dirac-equation for particle and anti-particle by flipping the sign of $M(\lambda,\rho)$, and they are given by,
\begin{eqnarray}
\Bigl(i\hbar\gamma^{\mu}\nabla_{\mu}-m+M(\lambda,\rho)\Bigr){\Psi}=0 \\
\bar{\Psi}\Bigl(i\hbar\gamma^{\mu}\nabla_{\mu}+m-M(\lambda,\rho)\Bigr)=0 
\end{eqnarray}

\section{Proposal for a neutrino equation}

Taking $m=0$, the neutrino equation can be proposed as,

\begin{eqnarray}
\Bigl(i\hbar\gamma^{\mu}\nabla_{\mu}-M(\lambda,\rho)\Bigr)\Psi=0 \label{WFeqnVacEminus}
\end{eqnarray}

\begin{eqnarray}
\bar{\Psi}\Bigl(i\hbar\gamma^{\mu}\nabla_{\mu}+M(\lambda,\rho)\Bigr)=0 \label{WFeqnVacEplus} 
\end{eqnarray}
where $M(\lambda,\rho)= \pm \lim_{m\to 0}\Bigl(\frac{m(1-Q)\lambda}{\rho}-\frac{\hbar^2}{2m\rho}\Box\lambda\Bigr)^{1/2}$.
\begin{eqnarray}
\lambda=\lim_{m\to 0} \frac{\hbar^2}{m^2(1-Q)}\nabla_{\mu}\Bigl(\lambda\frac{\nabla^{\mu}\sqrt{\rho}}{\sqrt{\rho}} \Bigr) \label{LambdaNiceEq2}
\end{eqnarray}
Its is to be noted that, even in the massless fermion case ($m=0$) the equations give a mass like contribution from the vacuum interaction term $M(\lambda,\rho)$ if the limiting quantity is non-zero.
This interesting mass contribution is not a constant, but a function of vacuum field $\lambda$ and quantum mechanical density $\rho$. Hence it can be real or complex depending upon the dynamical behavior of $\lambda$ and $\rho$. 
Taking into account quantum gravity correction, the extra term appears $M(\lambda,\rho)$ is no-longer a constant, 
in some specific scenarios it may behave like a constant. One can think about vacuum-mass as a quantity depends on 
space-time coordinate and even the rest-mass is a function of space-time coordinate. 
This implies an interesting possibility that, we might be
able to modify the mass of an object by properly adjusting the vacuum interaction. 
A.Sakharov had suggested that the gravitation might be electromagnetic phenomenon induced by the presence of matter in the quantum vacuum~\cite{Sakharov1968}. In that perspective, gravity cannot be considered as a fundamental interaction, it must arise from the interaction of the matter field with the quantum vacuum in the accelerating frames.

Bohmian quantum gravity reveals the nature of inertia in a more interesting manner along the similar lines of thought
as introduced in the quantum vacuum inertia hypothesis ~\cite{GrvaccInertia}. It was H. E. Puthoff  
who suggested  that the quantum vacuum might be responsible for inertia ~\cite{PuthoffZPEFluct1989}.
Later A. Rueda, B. Haisch and H. E. Puthoff indicated that the origin of inertial reaction forces can be explained as a zero-point field Lorentz force~\cite{InertiaZPFLorentz}.
It is proposed that the interaction of electrically charged elementary particles with the vacuum electromagnetic zero-point
field results into inertial forces. These ideas needs to be understood in the usual quantum field theory context. Results presented here will shed light on these kinds of theoretical extensions of quantum field theory. In the usual quantum field theory mass is just considered as a constant while taking into account the quantum-gravity idea, we need to extend the concept of mass from a constant to a quantity which depends on the vacuum field variable $\lambda$ and quantum mechanical probability density $\rho$. The vacuum interaction is not restricted to charged particles only, it can appear for any matter and the
interaction happens through the probability density $\rho$ of the particle with the vacuum field $\lambda$. One should distinguish the mass fluctuation due to the conformal factor $\Omega^2$ and the vacuum mass contributions appears in the theory through $\lambda$.

\section{Conclusion}

In this manuscript, a more general Dirac equation is derived using Bohmian Quantum Gravity consideration.
It is found that the Dirac equation contains an effective mass contribution arising from the vacuum. 
This vacuum mass term persist even for massless fermions, hence the equation might be useful to explain some of the mysterious properties of neutrinos. Dirac equation is coupled with a scalar vacuum field equation $\lambda$ via the particle density $\bar{\Psi}\Psi$. The appearance of vacuum-mass term to be taken as an important feature arising from the coupling of scalar vacuum field $\lambda$ with fermionic matter field.


\begin{thebibliography}{36}%
\makeatletter
\providecommand \@ifxundefined [1]{%
 \@ifx{#1\undefined}
}%
\providecommand \@ifnum [1]{%
 \ifnum #1\expandafter \@firstoftwo
 \else \expandafter \@secondoftwo
 \fi
}%
\providecommand \@ifx [1]{%
 \ifx #1\expandafter \@firstoftwo
 \else \expandafter \@secondoftwo
 \fi
}%
\providecommand \natexlab [1]{#1}%
\providecommand \enquote  [1]{``#1''}%
\providecommand \bibnamefont  [1]{#1}%
\providecommand \bibfnamefont [1]{#1}%
\providecommand \citenamefont [1]{#1}%
\providecommand \href@noop [0]{\@secondoftwo}%
\providecommand \href [0]{\begingroup \@sanitize@url \@href}%
\providecommand \@href[1]{\@@startlink{#1}\@@href}%
\providecommand \@@href[1]{\endgroup#1\@@endlink}%
\providecommand \@sanitize@url [0]{\catcode `\\12\catcode `\$12\catcode
  `\&12\catcode `\#12\catcode `\^12\catcode `\_12\catcode `\%12\relax}%
\providecommand \@@startlink[1]{}%
\providecommand \@@endlink[0]{}%
\providecommand \url  [0]{\begingroup\@sanitize@url \@url }%
\providecommand \@url [1]{\endgroup\@href {#1}{\urlprefix }}%
\providecommand \urlprefix  [0]{URL }%
\providecommand \Eprint [0]{\href }%
\providecommand \doibase [0]{http://dx.doi.org/}%
\providecommand \selectlanguage [0]{\@gobble}%
\providecommand \bibinfo  [0]{\@secondoftwo}%
\providecommand \bibfield  [0]{\@secondoftwo}%
\providecommand \translation [1]{[#1]}%
\providecommand \BibitemOpen [0]{}%
\providecommand \bibitemStop [0]{}%
\providecommand \bibitemNoStop [0]{.\EOS\space}%
\providecommand \EOS [0]{\spacefactor3000\relax}%
\providecommand \BibitemShut  [1]{\csname bibitem#1\endcsname}%
\let\auto@bib@innerbib\@empty
\bibitem [{\citenamefont {Shojai}\ and\ \citenamefont
  {Golshani}(1998)}]{Shojai_Article}%
  \BibitemOpen
  \bibfield  {author} {\bibinfo {author} {\bibfnamefont {F.}~\bibnamefont
  {Shojai}}\ and\ \bibinfo {author} {\bibfnamefont {M.}~\bibnamefont
  {Golshani}},\ }\href {\doibase 10.1142/S0217751X98000305} {\bibfield
  {journal} {\bibinfo  {journal} {Int. J. Mod. Phys. A}\ }\textbf {\bibinfo
  {volume} {13}},\ \bibinfo {pages} {677} (\bibinfo {year} {1998})}\BibitemShut
  {NoStop}%
\bibitem [{\citenamefont {{Shojai}}\ and\ \citenamefont
  {{Shojai}}(2004)}]{QMGeometry}%
  \BibitemOpen
  \bibfield  {author} {\bibinfo {author} {\bibfnamefont {F.}~\bibnamefont
  {{Shojai}}}\ and\ \bibinfo {author} {\bibfnamefont {A.}~\bibnamefont
  {{Shojai}}},\ }\href@noop {} {\bibfield  {journal} {\bibinfo  {journal}
  {ArXiv e-prints}\ } (\bibinfo {year} {2004})},\ \Eprint
  {http://arxiv.org/abs/gr-qc/0404102} {gr-qc/0404102} \BibitemShut {NoStop}%
\bibitem [{\citenamefont {{Gabay}}\ and\ \citenamefont
  {{Joseph}}(2018{\natexlab{a}})}]{GabayJoseph1}%
  \BibitemOpen
  \bibfield  {author} {\bibinfo {author} {\bibfnamefont {D.}~\bibnamefont
  {{Gabay}}}\ and\ \bibinfo {author} {\bibfnamefont {S.~K.}\ \bibnamefont
  {{Joseph}}},\ }\href@noop {} {\bibfield  {journal} {\bibinfo  {journal}
  {ArXiv e-prints}\ } (\bibinfo {year} {2018}{\natexlab{a}})},\ \Eprint
  {http://arxiv.org/abs/1801.00161} {arXiv:1801.00161 [gr-qc]} \BibitemShut
  {NoStop}%
\bibitem [{\citenamefont {{Gabay}}\ and\ \citenamefont
  {{Joseph}}(2018{\natexlab{b}})}]{GabayJoseph2}%
  \BibitemOpen
  \bibfield  {author} {\bibinfo {author} {\bibfnamefont {D.}~\bibnamefont
  {{Gabay}}}\ and\ \bibinfo {author} {\bibfnamefont {S.~K.}\ \bibnamefont
  {{Joseph}}},\ }\href@noop {} {\bibfield  {journal} {\bibinfo  {journal}
  {ArXiv e-prints}\ } (\bibinfo {year} {2018}{\natexlab{b}})},\ \Eprint
  {http://arxiv.org/abs/1802.07678} {arXiv:1802.07678 [gr-qc]} \BibitemShut
  {NoStop}%
\bibitem [{\citenamefont {Nottale}(1993)}]{NottaleBook1}%
  \BibitemOpen
  \bibfield  {author} {\bibinfo {author} {\bibfnamefont {L.}~\bibnamefont
  {Nottale}},\ }\href {\doibase 10.1142/1579} {\emph {\bibinfo {title} {Fractal
  Space-Time and Microphysics}}}\ (\bibinfo  {publisher} {World Scientific},\
  \bibinfo {year} {1993})\BibitemShut {NoStop}%
\bibitem [{\citenamefont {Nottale}(2011)}]{NottaleBook2}%
  \BibitemOpen
  \bibfield  {author} {\bibinfo {author} {\bibfnamefont {L.}~\bibnamefont
  {Nottale}},\ }\href {\doibase 10.1142/p752} {\emph {\bibinfo {title} {Scale
  Relativity and Fractal Space-Time}}}\ (\bibinfo  {publisher} {Imperial
  College Press},\ \bibinfo {year} {2011})\BibitemShut {NoStop}%
\bibitem [{\citenamefont {Capozziello}\ and\ \citenamefont
  {Faraoni}(2011)}]{BeyondEinsteinGravity}%
  \BibitemOpen
  \bibfield  {author} {\bibinfo {author} {\bibfnamefont {S.}~\bibnamefont
  {Capozziello}}\ and\ \bibinfo {author} {\bibfnamefont {V.}~\bibnamefont
  {Faraoni}},\ }\href@noop {} {\emph {\bibinfo {title} {Beyond Einstein
  Gravity}}},\ \bibinfo {edition} {first edition}\ ed.\ (\bibinfo  {publisher}
  {Springer Netherlands},\ \bibinfo {address} {Dordrecht, Holland},\ \bibinfo
  {year} {2011})\BibitemShut {NoStop}%
\bibitem [{\citenamefont {{Scholz}}(2017)}]{WeylReview2017}%
  \BibitemOpen
  \bibfield  {author} {\bibinfo {author} {\bibfnamefont {E.}~\bibnamefont
  {{Scholz}}},\ }\href@noop {} {\bibfield  {journal} {\bibinfo  {journal}
  {ArXiv e-prints}\ } (\bibinfo {year} {2017})},\ \Eprint
  {http://arxiv.org/abs/1703.03187} {arXiv:1703.03187 [math.HO]} \BibitemShut
  {NoStop}%
\bibitem [{\citenamefont {Bergmann}(1968)}]{Bergmann1968}%
  \BibitemOpen
  \bibfield  {author} {\bibinfo {author} {\bibfnamefont {P.~G.}\ \bibnamefont
  {Bergmann}},\ }\href {\doibase 10.1007/BF00668828} {\bibfield  {journal}
  {\bibinfo  {journal} {Int. J. Theor. Phys.}\ }\textbf {\bibinfo {volume}
  {1}},\ \bibinfo {pages} {25} (\bibinfo {year} {1968})}\BibitemShut {NoStop}%
\bibitem [{\citenamefont {Deruelle}\ \emph {et~al.}(2010)\citenamefont
  {Deruelle}, \citenamefont {Sasaki}, \citenamefont {Sendouda},\ and\
  \citenamefont {Yamauchi}}]{fR_HamFormlation}%
  \BibitemOpen
  \bibfield  {author} {\bibinfo {author} {\bibfnamefont {N.}~\bibnamefont
  {Deruelle}}, \bibinfo {author} {\bibfnamefont {M.}~\bibnamefont {Sasaki}},
  \bibinfo {author} {\bibfnamefont {Y.}~\bibnamefont {Sendouda}}, \ and\
  \bibinfo {author} {\bibfnamefont {D.}~\bibnamefont {Yamauchi}},\ }\href
  {\doibase 10.1143/PTP.123.169} {\bibfield  {journal} {\bibinfo  {journal}
  {Progress. Theor. Phys.}\ }\textbf {\bibinfo {volume} {123}},\ \bibinfo
  {pages} {169} (\bibinfo {year} {2010})}\BibitemShut {NoStop}%
\bibitem [{\citenamefont {Ferraro}\ and\ \citenamefont
  {Fiorini}(2007)}]{fofT1}%
  \BibitemOpen
  \bibfield  {author} {\bibinfo {author} {\bibfnamefont {R.}~\bibnamefont
  {Ferraro}}\ and\ \bibinfo {author} {\bibfnamefont {F.}~\bibnamefont
  {Fiorini}},\ }\href {\doibase 10.1103/PhysRevD.75.084031} {\bibfield
  {journal} {\bibinfo  {journal} {Phys. Rev. D}\ }\textbf {\bibinfo {volume}
  {75}},\ \bibinfo {pages} {084031} (\bibinfo {year} {2007})}\BibitemShut
  {NoStop}%
\bibitem [{\citenamefont {Ferraro}\ and\ \citenamefont
  {Fiorini}(2008)}]{fofT2}%
  \BibitemOpen
  \bibfield  {author} {\bibinfo {author} {\bibfnamefont {R.}~\bibnamefont
  {Ferraro}}\ and\ \bibinfo {author} {\bibfnamefont {F.}~\bibnamefont
  {Fiorini}},\ }\href {\doibase 10.1103/PhysRevD.78.124019} {\bibfield
  {journal} {\bibinfo  {journal} {Phys. Rev. D}\ }\textbf {\bibinfo {volume}
  {78}},\ \bibinfo {pages} {124019} (\bibinfo {year} {2008})}\BibitemShut
  {NoStop}%
\bibitem [{\citenamefont {Bekenstein}(2011)}]{Bekenstein2011_TeVeS}%
  \BibitemOpen
  \bibfield  {author} {\bibinfo {author} {\bibfnamefont {J.~D.}\ \bibnamefont
  {Bekenstein}},\ }\href {\doibase 10.1098/rsta.2011.0282} {\bibfield
  {journal} {\bibinfo  {journal} {Philos. Trans. Royal Soc. A}\ }\textbf
  {\bibinfo {volume} {369}},\ \bibinfo {pages} {5003} (\bibinfo {year}
  {2011})}\BibitemShut {NoStop}%
\bibitem [{\citenamefont {Bekenstein}(2004)}]{BekensteinPRD_TeVeS}%
  \BibitemOpen
  \bibfield  {author} {\bibinfo {author} {\bibfnamefont {J.~D.}\ \bibnamefont
  {Bekenstein}},\ }\href {\doibase 10.1103/PhysRevD.70.083509} {\bibfield
  {journal} {\bibinfo  {journal} {Phys. Rev. D}\ }\textbf {\bibinfo {volume}
  {70}},\ \bibinfo {pages} {083509} (\bibinfo {year} {2004})}\BibitemShut
  {NoStop}%
\bibitem [{\citenamefont {Moffat}(2006)}]{Moffat_STVG}%
  \BibitemOpen
  \bibfield  {author} {\bibinfo {author} {\bibfnamefont {J.~W.}\ \bibnamefont
  {Moffat}},\ }\href {http://stacks.iop.org/1475-7516/2006/i=03/a=004}
  {\bibfield  {journal} {\bibinfo  {journal} {‎J. Cosmol. Astropart. Phys}\
  ,\ \bibinfo {pages} {004}} (\bibinfo {year} {2006})}\BibitemShut {NoStop}%
\bibitem [{\citenamefont {Moffat}\ and\ \citenamefont
  {Toth}(2009)}]{MOG_Moffat}%
  \BibitemOpen
  \bibfield  {author} {\bibinfo {author} {\bibfnamefont {J.~W.}\ \bibnamefont
  {Moffat}}\ and\ \bibinfo {author} {\bibfnamefont {V.~T.}\ \bibnamefont
  {Toth}},\ }\href@noop {} {\bibfield  {journal} {\bibinfo  {journal} {‎Mon.
  Notices Royal Astron. Soc}\ }\textbf {\bibinfo {volume} {395}},\ \bibinfo
  {pages} {L25} (\bibinfo {year} {2009})}\BibitemShut {NoStop}%
\bibitem [{\citenamefont {Bohm}\ and\ \citenamefont {Hiley}(1975)}]{Bohm1975}%
  \BibitemOpen
  \bibfield  {author} {\bibinfo {author} {\bibfnamefont {D.~J.}\ \bibnamefont
  {Bohm}}\ and\ \bibinfo {author} {\bibfnamefont {B.~J.}\ \bibnamefont
  {Hiley}},\ }\href@noop {} {\bibfield  {journal} {\bibinfo  {journal} {Found.
  Phys.}\ }\textbf {\bibinfo {volume} {5}},\ \bibinfo {pages} {93} (\bibinfo
  {year} {1975})}\BibitemShut {NoStop}%
\bibitem [{\citenamefont {Bohm}(1952{\natexlab{a}})}]{BohmI}%
  \BibitemOpen
  \bibfield  {author} {\bibinfo {author} {\bibfnamefont {D.}~\bibnamefont
  {Bohm}},\ }\href {\doibase 10.1103/PhysRev.85.166} {\bibfield  {journal}
  {\bibinfo  {journal} {Phys. Rev.}\ }\textbf {\bibinfo {volume} {85}},\
  \bibinfo {pages} {166} (\bibinfo {year} {1952}{\natexlab{a}})}\BibitemShut
  {NoStop}%
\bibitem [{\citenamefont {Bohm}(1952{\natexlab{b}})}]{BohmII}%
  \BibitemOpen
  \bibfield  {author} {\bibinfo {author} {\bibfnamefont {D.}~\bibnamefont
  {Bohm}},\ }\href {\doibase 10.1103/PhysRev.85.180} {\bibfield  {journal}
  {\bibinfo  {journal} {Phys. Rev.}\ }\textbf {\bibinfo {volume} {85}},\
  \bibinfo {pages} {180} (\bibinfo {year} {1952}{\natexlab{b}})}\BibitemShut
  {NoStop}%
\bibitem [{\citenamefont {Holland}(1995)}]{PHollandBook}%
  \BibitemOpen
  \bibfield  {author} {\bibinfo {author} {\bibfnamefont {P.~R.}\ \bibnamefont
  {Holland}},\ }\href@noop {} {\emph {\bibinfo {title} {The Quantum Theory of
  Motion}}},\ \bibinfo {edition} {first edition}\ ed.\ (\bibinfo  {publisher}
  {Cambridge University Press},\ \bibinfo {address} {Cambridge, United
  Kingdom},\ \bibinfo {year} {1995})\BibitemShut {NoStop}%
\bibitem [{\citenamefont {{Goldstein}}(1995)}]{SheldonReview}%
  \BibitemOpen
  \bibfield  {author} {\bibinfo {author} {\bibfnamefont {S.}~\bibnamefont
  {{Goldstein}}},\ }\href@noop {} {\bibfield  {journal} {\bibinfo  {journal}
  {eprint arXiv:quant-ph/9512027}\ } (\bibinfo {year} {1995})},\ \Eprint
  {http://arxiv.org/abs/quant-ph/9512027} {quant-ph/9512027} \BibitemShut
  {NoStop}%
\bibitem [{\citenamefont {Shojai}\ and\ \citenamefont
  {Shojai}(2004)}]{ShojaiCnstrAlgebra}%
  \BibitemOpen
  \bibfield  {author} {\bibinfo {author} {\bibfnamefont {A.}~\bibnamefont
  {Shojai}}\ and\ \bibinfo {author} {\bibfnamefont {F.}~\bibnamefont
  {Shojai}},\ }\href {http://stacks.iop.org/0264-9381/21/i=1/a=001} {\bibfield
  {journal} {\bibinfo  {journal} {Class. Quantum Grav.}\ }\textbf {\bibinfo
  {volume} {21}},\ \bibinfo {pages} {1} (\bibinfo {year} {2004})},\ \Eprint
  {http://arxiv.org/abs/gr-qc/0311076} {gr-qc/0311076} \BibitemShut {NoStop}%
\bibitem [{\citenamefont {Shojai}\ and\ \citenamefont
  {Shojai}(2001)}]{ShojaiBohmianQM}%
  \BibitemOpen
  \bibfield  {author} {\bibinfo {author} {\bibfnamefont {A.}~\bibnamefont
  {Shojai}}\ and\ \bibinfo {author} {\bibfnamefont {F.}~\bibnamefont
  {Shojai}},\ }\href {http://stacks.iop.org/1402-4896/64/i=5/a=003} {\bibfield
  {journal} {\bibinfo  {journal} {Phys. Scripta}\ }\textbf {\bibinfo {volume}
  {64}},\ \bibinfo {pages} {413} (\bibinfo {year} {2001})},\ \Eprint
  {http://arxiv.org/abs/quant-ph/0109025} {quant-ph/0109025} \BibitemShut
  {NoStop}%
\bibitem [{\citenamefont {Shojai}\ and\ \citenamefont
  {Shojai}(2000)}]{Shojai_ScalarTensor}%
  \BibitemOpen
  \bibfield  {author} {\bibinfo {author} {\bibfnamefont {F.}~\bibnamefont
  {Shojai}}\ and\ \bibinfo {author} {\bibfnamefont {A.}~\bibnamefont
  {Shojai}},\ }\href {\doibase 10.1142/S0217751X0000080X} {\bibfield  {journal}
  {\bibinfo  {journal} {Int. J. Mod. Phys. A}\ }\textbf {\bibinfo {volume}
  {15}},\ \bibinfo {pages} {1859} (\bibinfo {year} {2000})},\ \Eprint
  {http://arxiv.org/abs/gr-qc/0010012} {gr-qc/0010012} \BibitemShut {NoStop}%
\bibitem [{\citenamefont {Thiemann}(2007)}]{Thiemann_2007}%
  \BibitemOpen
  \bibfield  {author} {\bibinfo {author} {\bibfnamefont {T.}~\bibnamefont
  {Thiemann}},\ }\href@noop {} {\emph {\bibinfo {title} {Modern Canonical
  Quantum General Relativity}}},\ Cambridge Monographs on Mathematical Physics\
  (\bibinfo  {publisher} {Cambridge University Press},\ \bibinfo {year}
  {2007})\BibitemShut {NoStop}%
\bibitem [{\citenamefont {Rovelli}\ and\ \citenamefont
  {Vidotto}(2014)}]{rovelli_vidotto_2014}%
  \BibitemOpen
  \bibfield  {author} {\bibinfo {author} {\bibfnamefont {C.}~\bibnamefont
  {Rovelli}}\ and\ \bibinfo {author} {\bibfnamefont {F.}~\bibnamefont
  {Vidotto}},\ }\href@noop {} {\emph {\bibinfo {title} {Covariant Loop Quantum
  Gravity}}}\ (\bibinfo  {publisher} {Cambridge University Press},\ \bibinfo
  {year} {2014})\BibitemShut {NoStop}%
\bibitem [{\citenamefont {Gambini}\ and\ \citenamefont
  {Pullin}(2011)}]{Gambini2011}%
  \BibitemOpen
  \bibfield  {author} {\bibinfo {author} {\bibfnamefont {R.}~\bibnamefont
  {Gambini}}\ and\ \bibinfo {author} {\bibfnamefont {J.}~\bibnamefont
  {Pullin}},\ }\href@noop {} {\emph {\bibinfo {title} {{A first course in loop
  quantum gravity}}}}\ (\bibinfo {year} {2011})\BibitemShut {NoStop}%
\bibitem [{\citenamefont {Ashtekar}\ and\ \citenamefont
  {Pullin}(2017)}]{AshtekarBook2017}%
  \BibitemOpen
  \bibfield  {author} {\bibinfo {author} {\bibfnamefont {A.}~\bibnamefont
  {Ashtekar}}\ and\ \bibinfo {author} {\bibfnamefont {J.}~\bibnamefont
  {Pullin}},\ }\href {\doibase 10.1142/10445} {\emph {\bibinfo {title} {Loop
  Quantum Gravity}}}\ (\bibinfo  {publisher} {World Scientific},\ \bibinfo
  {year} {2017})\BibitemShut {NoStop}%
\bibitem [{\citenamefont {Nelson}(1985)}]{NelsonBook}%
  \BibitemOpen
  \bibfield  {author} {\bibinfo {author} {\bibfnamefont {E.}~\bibnamefont
  {Nelson}},\ }\href@noop {} {\emph {\bibinfo {title} {Quantum
  Fluctuations}}},\ Princeton series in physics\ (\bibinfo  {publisher}
  {Princeton, N.J. : Princeton University Press},\ \bibinfo {year}
  {1985})\BibitemShut {NoStop}%
\bibitem [{\citenamefont {Roy}(1992)}]{Sisir_Roy1}%
  \BibitemOpen
  \bibfield  {author} {\bibinfo {author} {\bibfnamefont {S.}~\bibnamefont
  {Roy}},\ }\href@noop {} {\bibfield  {journal} {\bibinfo  {journal} {Acta.
  Appl. Math.}\ }\textbf {\bibinfo {volume} {26}},\ \bibinfo {pages} {209}
  (\bibinfo {year} {1992})}\BibitemShut {NoStop}%
\bibitem [{\citenamefont {Asanov}\ \emph {et~al.}()\citenamefont {Asanov},
  \citenamefont {Ponomarenko},\ and\ \citenamefont {Roy}}]{Sisir_Roy2}%
  \BibitemOpen
  \bibfield  {author} {\bibinfo {author} {\bibfnamefont {G.~S.}\ \bibnamefont
  {Asanov}}, \bibinfo {author} {\bibfnamefont {S.~P.}\ \bibnamefont
  {Ponomarenko}}, \ and\ \bibinfo {author} {\bibfnamefont {S.}~\bibnamefont
  {Roy}},\ }\href@noop {} {\bibfield  {journal} {\bibinfo  {journal}
  {Fortschritte der Physik}\ }\textbf {\bibinfo {volume} {36}},\ \bibinfo
  {pages} {679}}\BibitemShut {NoStop}%
\bibitem [{\citenamefont {Smolin}(1986)}]{Smolin_QFluct1}%
  \BibitemOpen
  \bibfield  {author} {\bibinfo {author} {\bibfnamefont {L.}~\bibnamefont
  {Smolin}},\ }\href {http://stacks.iop.org/0264-9381/3/i=3/a=009} {\bibfield
  {journal} {\bibinfo  {journal} {Classical and Quantum Gravity}\ }\textbf
  {\bibinfo {volume} {3}},\ \bibinfo {pages} {347} (\bibinfo {year}
  {1986})}\BibitemShut {NoStop}%
\bibitem [{\citenamefont {{Sakharov}}(1968)}]{Sakharov1968}%
  \BibitemOpen
  \bibfield  {author} {\bibinfo {author} {\bibfnamefont {A.~D.}\ \bibnamefont
  {{Sakharov}}},\ }\href@noop {} {\bibfield  {journal} {\bibinfo  {journal}
  {Sov. Phys. Dokl.}\ }\textbf {\bibinfo {volume} {12}},\ \bibinfo {pages}
  {1040} (\bibinfo {year} {1968})}\BibitemShut {NoStop}%
\bibitem [{\citenamefont {Rueda}\ and\ \citenamefont
  {Haisch}(2005)}]{GrvaccInertia}%
  \BibitemOpen
  \bibfield  {author} {\bibinfo {author} {\bibfnamefont {A.}~\bibnamefont
  {Rueda}}\ and\ \bibinfo {author} {\bibfnamefont {B.}~\bibnamefont {Haisch}},\
  }\href {\doibase 10.1002/andp.200510147} {\bibfield  {journal} {\bibinfo
  {journal} {Annalen der Physik}\ }\textbf {\bibinfo {volume} {14}},\ \bibinfo
  {pages} {479} (\bibinfo {year} {2005})}\BibitemShut {NoStop}%
\bibitem [{\citenamefont {Puthoff}(1989)}]{PuthoffZPEFluct1989}%
  \BibitemOpen
  \bibfield  {author} {\bibinfo {author} {\bibfnamefont {H.~E.}\ \bibnamefont
  {Puthoff}},\ }\href@noop {} {\bibfield  {journal} {\bibinfo  {journal} {Phys.
  Rev. A}\ }\textbf {\bibinfo {volume} {39}},\ \bibinfo {pages} {2333}
  (\bibinfo {year} {1989})}\BibitemShut {NoStop}%
\bibitem [{\citenamefont {Rueda}\ \emph {et~al.}(1994)\citenamefont {Rueda},
  \citenamefont {Haisch},\ and\ \citenamefont {Puthoff}}]{InertiaZPFLorentz}%
  \BibitemOpen
  \bibfield  {author} {\bibinfo {author} {\bibfnamefont {A.}~\bibnamefont
  {Rueda}}, \bibinfo {author} {\bibfnamefont {B.}~\bibnamefont {Haisch}}, \
  and\ \bibinfo {author} {\bibfnamefont {H.~E.}\ \bibnamefont {Puthoff}},\
  }\href@noop {} {\bibfield  {journal} {\bibinfo  {journal} {Phys. Rev. A}\
  }\textbf {\bibinfo {volume} {49}},\ \bibinfo {pages} {678} (\bibinfo {year}
  {1994})}\BibitemShut {NoStop}%
\end{thebibliography}

%

\end{document}